\documentstyle[12pt]{article}
 \advance\oddsidemargin by -1in
\advance\evensidemargin
 by -1in
\advance\topmargin by -0.0in
\makeindex

\newcommand\beq{\begin{equation}}
 \newcommand\eeq{\end{equation}}
\newcommand\beqn{\begin{eqnarray}}
 \newcommand\eeqn{\end{eqnarray}}

\begin{document}
\hspace*{9cm}{\Large LPTHE96-01}

\vskip 2 truemm
\noindent\hspace*{9.6cm}{\Large MPIH-V07-1996}\\

\vspace*{4cm}

\centerline{\Large\bf Novel Mechanism of Nucleon Stopping}\medskip
\centerline{\Large\bf in Heavy Ion Collisions}
\vbox to 2 truecm {}

\centerline{\large\bf A. Capella}\smallskip
\centerline{Laboratoire de Physique Th\'eorique et Hautes Energies\footnote{Laboratoire
associ\'e au Centre National de la Recherche Scientifique - URA D0063}}
\centerline{Universit\'e de Paris XI, b\^atiment 211, 91405 Orsay Cedex, France}  \medskip
\medskip 
\centerline{{\large\bf B.Z.~Kopeliovich}\footnote{On leave from Joint Institute for
 Nuclear Research, Laboratory of Nuclear Problems,
\newline 
\hspace*{0.5cm} Dubna, 141980 Moscow Region, Russia.}}\smallskip
\centerline{Max-Planck Institut f\"ur Kernphysik, Postfach 103980}
\centerline{69029 Heidelberg, Germany}

\bigskip\bigskip
\baselineskip=20pt 
\noindent 
{\bf Abstract} \par 
When a diquark does not fragment directly but breaks in such a way that only one of its
quarks gets into the produced baryon, the latter is produced closer to mid rapidities.
The relative size of this diquark breaking component increases quite fast with
increasing energy. We show that at a given energy it also increases with the atomic mass
number and with the centrality of the collision and that it allows to explain the
rapidity distribution of the net baryon number ($p$-$\bar{p}$) in $SS$ central
collisions. Predictions for $Pb$-$Pb$ collisions are presented. \par

\vbox to 2cm{}

\newpage
In most string models particle production in a $pp$ collision occurs mainly in the form
of two diquark-quark strings. There is strong experimental evidence \cite{[1]} that with a
large probability the diquark acts as a single entity and fragments directly into
a leading baryon. Since the diquark is fast in average, taking a large fraction of the
incoming proton momentum, the produced baryon will be in the proton fragmentation region.
Moreover, it is easy to see that the corresponding baryon spectrum at $y^* \sim 0$
decreases, quite fast with energy - roughly as $s^{-1}$ \cite{[1]}-\cite{[3]}. 
A diquark breaking
component in which only one of two quarks gets into the produced baryon is of course
present. This component produces baryons at smaller values of $y^*$ and the produced
baryon distribution at $y^* \sim 0$ only decreases like $s^{-1/4}$ \cite{[3]}. In the Dual Parton
Model (DPM) and Quark Gluon String Model (QGSM) such a component is not explicitly taken
into account \cite{[2]}, \cite{[4]}. However, it has been shown by Kopeliovich and 
Zakharov \cite{[3]} that the
energy dependence of $d\sigma/dy$ at $y^* = 0$ for the difference $p$ minus $\bar{p}$
measured by two different collaborations \cite{[5]}, \cite{[6]} 
in the ISR energy range can only be
understood with a diquark breaking component of a relative size of about 20 $\%$ which
has to be introduced in the string models as a separate component. The authors of ref.
\cite{[3]} have also proposed an interpretation of the diquark breaking component as a result
of a color exchange which converts the diquark from a $\{\bar{3} \}$ to a $\{ 6\}$ color
state. While this is an attractive mechanism, it may not be the only one. If to assume
that the relative longitudinal momenta of the quarks inside the diquark may be 
substantially different, a diquark breaking with excitation of a high-mass state
could also result in the baryon production at mid rapidities.
\par

In the present work we do not specify the origin of the diquark breaking component and
do not attempt to compute its size theoretically (for a calculation in perturbative QCD
see ref. \cite{[3]}). Following the phenomenological analysis in ref. \cite{[3]} 
we determine the
latter from experiment. Also, we concentrate ourselves to present CERN energies of
$\sqrt{s} \sim$ 20 GeV and make only some comments about our expectations for RHIC or
LHC energies. We shall show that in heavy ion collisions the relative size of the
diquark breaking component increases with atomic mass number and with the centrality
of the collision. We show that in central $SS$ collision the rapidity distribution of
the difference $p$-$\bar{p}$ - which in DPM has a sharp dip at mid rapidities 
\cite{[4]} not present in the
data \cite{[7]} - is consistent with experiment after introducing the diquark breaking
component - with relative size determined from $pp$ data as explained above. The
corresponding distributions for central $Pb$-$Pb$ collisions at present CERN energies
can then be computed, and are given as a prediction of our diquark breaking mechanism.
\par

In DPM, the dominant component in $pp$ scattering is the two string configuration
depicted in fig. 1. Here the diquark fragments directly into a final state baryon. The
rapidity distribution $dN/dy$ of a hadron in a string is obtained from a convolution
of the momentum distribution and fragmentation functions \cite{[1]}. All formulas 
and parameters relative to baryon production in nucleon-nucleon collision 
are given in refs. \cite{[2]} and \cite{[4]}.
The corresponding results will be denoted $dN_{DP}/dy$ (where $DP$ stands for diquark
preserving). As mentioned above there is also the possibility that the diquark breaks
in the way depicted in fig. 2. In this case the baryon rapidity distribution follows
that of the valence quark and thus the baryon is slower than in the $DP$
component of fig. 1. We then have

\begin{figure}[tbh]
\includegraphics{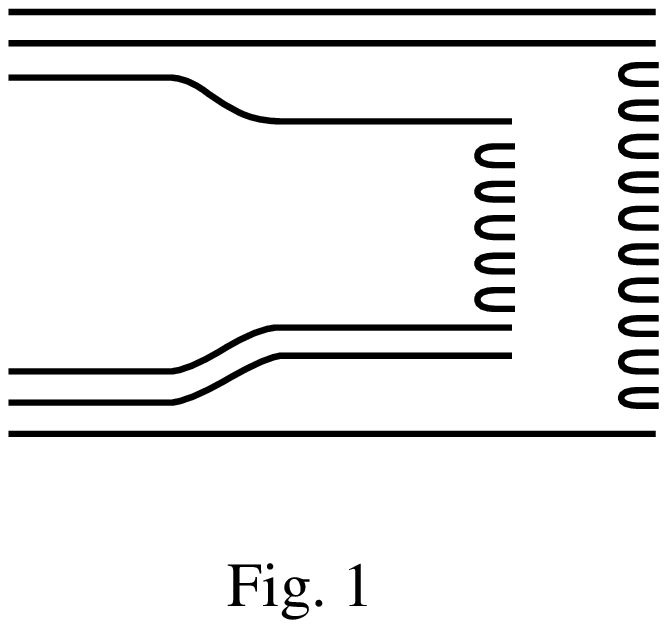}
\includegraphics{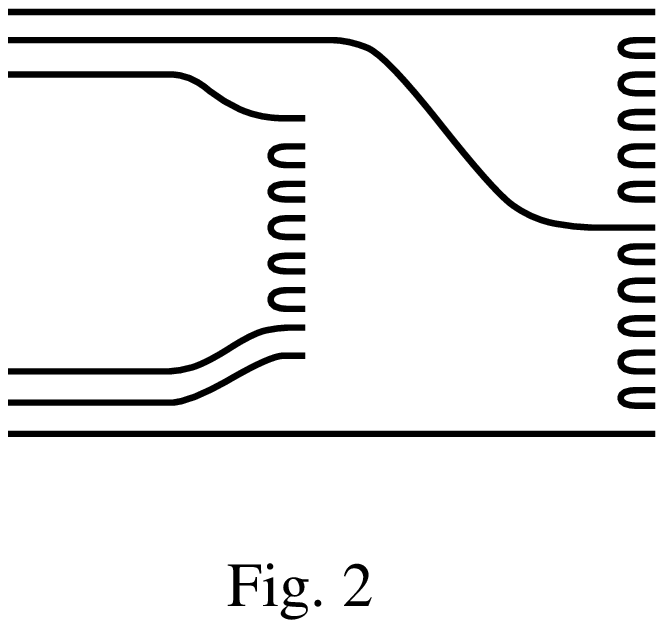}
\begin{center}
\vspace{6cm}
\parbox{13cm}
{\caption[Delta]
{Diquark preserving component in nucleon-nucleon scattering.}
\label{fig1}}
{\caption[Delta]
{Diquark breaking component in nucleon-nucleon scattering.}
\label{fig2}}
\end{center}
\end{figure}

$${dN_{DB} \over dy}(y) = {C \over 2} \left [ \widetilde{\rho}_{q_v}(y) +
\widetilde{\rho}_{q_v}(-y) \right ] \ \ \ , \eqno(1)$$

\noindent where $DB$ stands for diquark breaking and $\widetilde{\rho}_{q_v}(y)$ is the
valence quark rapidity distribution. This function and the value of the
constant $C$ are given below. Obviously, in the final formula, the $DB$ component has
to be weighted by the ratio $\sigma_{DB}/\sigma_{in}$ of the $DB$ over the inelastic
cross-section. \par

Let us now turn to a nucleon-nucleus ($NA$) collision. Here we can also have a $DP$ and
$DB$ components. In the case of a double inelastic collision they are depicted in figs. 3
and 4, respectively. In the $NA$ case (as well as in a $NN$ configuration involving
\begin{figure}[tbh]
\includegraphics{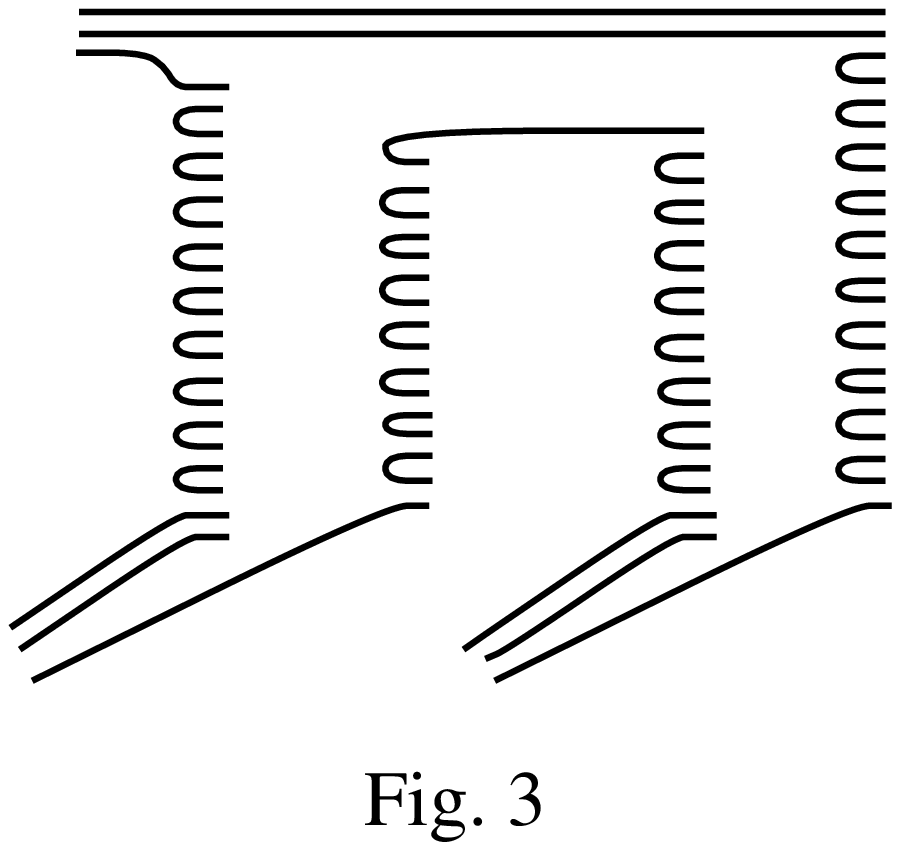}
\includegraphics{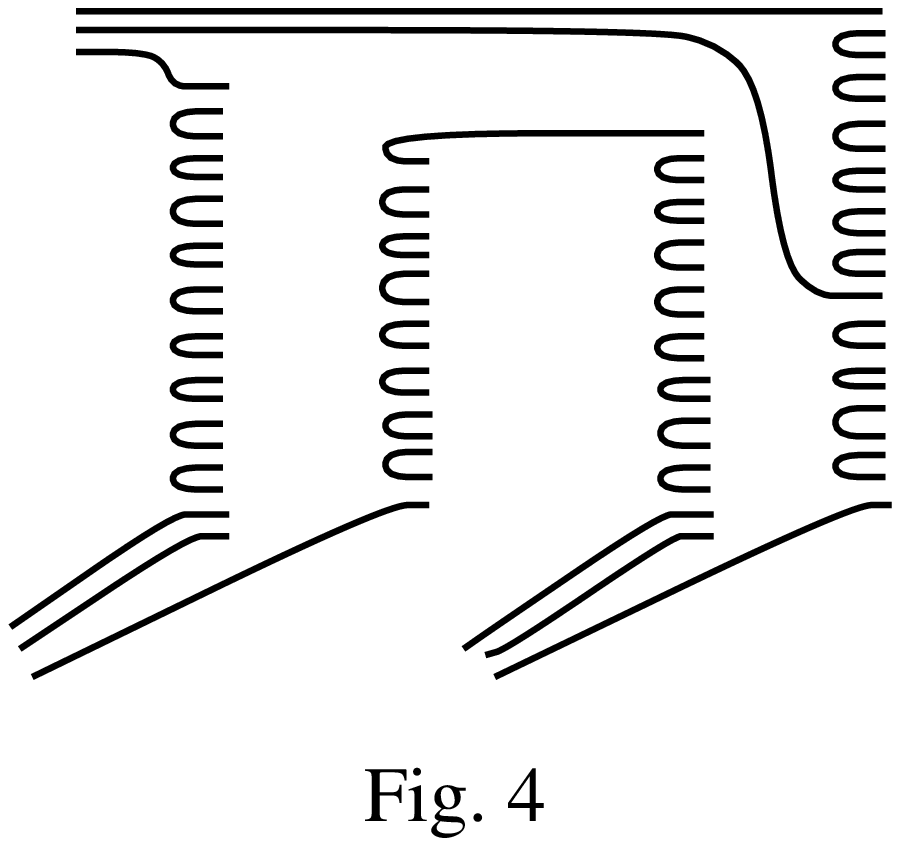}
\begin{center}
\vspace{6cm}
\parbox{13cm}
{\caption[Delta]
{Diquark preserving component for nucleon-nucleus scattering for two
inelastic collisions.}
\label{fig3}}
{\caption[Delta]
{Same as fig. 3 for the diquark breaking component.}
\label{fig4}}
\end{center}
\end{figure}
at least four strings) it is also possible to have a configuration as the one
depicted in fig. 5 in which the baryon number is associated to a gluon or sea quark
of the incoming nucleon. For the moment we disregard such a configuration but shall
come back to it later on. Let us now split the nucleon-nucleon ($NN$) cross-section
into its diquark preserving and diquark breaking components $\sigma_{in} = \sigma_{DP} +
\sigma_{DB}$. We assume that once the diquark has been destroyed in a collision
with one nucleon of the nucleus it cannot be reconstructed in further collisions
with other nucleons of the nucleus. The $NA$ cross-section involving $n$ inelastic
$NN$ collisions is then given by

\begin{figure}[b]
\includegraphics{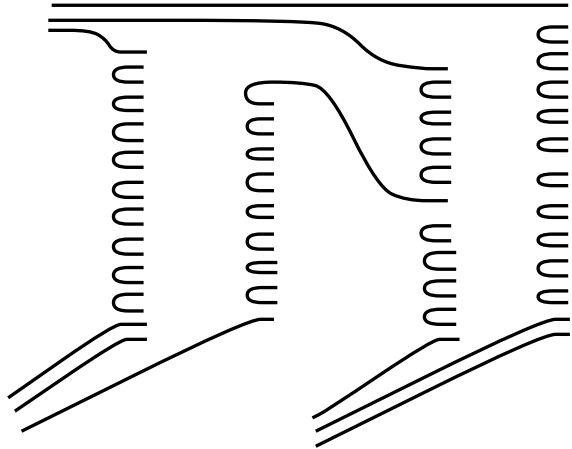}
\begin{center}
\vspace{6cm}
\parbox{13cm}
{\caption[Delta]
{Same as fig. 4 but here the produced baryon has the rapidity
distribution of a sea quark.}
\label{fig5}}
\end{center}
\end{figure}

$$\sigma_{DB,n}^{NA}(b) = {A \choose n} \sum_{i=1}^n {n \choose i} \sigma_{DB}^i \
\sigma_{DP}^{n-i} \ T_A^n(b) \left [ 1 - \sigma_{in} \ T_A(b) \right ]^{A-n} \ \ .
\eqno(2)$$~%

Here $T_A(b)$ is the standard nuclear profile function at impact parameter $b$,
normalized to unity. In (2) we have replaced the usual factor $\sigma_{in}^n \ T_A^n$
corresponding to the cross-section for $n$ inelastic collisions by the product
$\sum\limits_{i=1}^n {n \choose i} \sigma_{DB}^i \ \sigma_{DP}^{n-1} \ T_A^n =
(\sigma_{in}^n - \sigma_{DP}^n)T_A^n(b)$. Indeed only the term $\sigma_{DP}^n \ T_A^n$
will contribute to the diquark preserving cross-section. Summing in $n$ we have

$$\sigma_{DB}^{NA}(b) = \sum_{n=1}^A \sigma_{DB,n}^{NA}(b) = 1 - \left [ 1 - \sigma_{DB}
\ T_A(b) \right ]^A \ \ \ . \eqno(3)$$

\noindent Eq. (3) shows that the diquark preserving cross-section belongs to a class of
process \cite{[8]} which has only self-absorption (or self-shadowing). Obviously

$$\sigma_{DP}^{NA}(b) = \sigma_{in}^{NA}(b) - \sigma_{DB}^{NA}(b) \ \ \ . \eqno(4)$$

\noindent Since $\sigma_{DB} < \sigma_{in}$, it is clear from (3) and (4) that
$\sigma_{DB}^{NA}$ increases faster with $A$ than $\sigma_{DP}^{NA}$. Actually, when
$\sigma_{DB}$ is sufficiently small to neglect in (3) second and higher powers of
$\sigma_{DB}$, $\sigma_{DB}^{NA}$ will increase linearly with $A$. This proves the
result stated above that the relative size of the $DB$ component increases with
increasing $A$. The result can be easily generalized to an $AB$ collision. We
have \cite{[9]}

$$\sigma_{DB}^{AB}(b) = 1 - \left ( 1 - \sigma_{DB} \ T_{AB} (b) \right )^{AB}
\ \ \ . \eqno(5)$$

\noindent where $T_{AB}(\vec{b}) = \int d^2s \ T_A(\vec{s}) \ T_B(\vec{b} - \vec{s})$.
For $\sigma_{DB}$ sufficiently small we get from (5), after integration in impact
parameter, $\sigma_{DB}^{AB} = AB \ \sigma_{DB}$. \par

In order to compute the nucleon rapidity distribution we only need to specify the
values of $\sigma_{DB}$ as well as the value $C_{NN}$ of the constant $C$ in (1) for
an $NN$ collision, as well as the function $\widetilde{\rho}_{q_v}^n(y) \equiv Z \
\rho_{q_v}^n(Z)$ in (1). Here $Z = e^{-\Delta y}$, $\Delta y = y - Y_{MAX}$. The
function $\rho_{q_v}^n(Z)$ is explicitly given in DPM and QGSM and depends on the
number $n$ of inelastic collisions suffered by the nucleon ($n = 1$ in Figs.~1 and 2
and $n = 2$ in Figs.~3-5). Indeed, in these models the nucleon splits into $2n$
components (valence quark, diquark and $n - 1$ gluons or $q_s$-$\bar{q}_s$ pairs) and
the joint moment distribution is written as a product of the $2n$ factors controlling the
$Z \sim 0$ behaviour of the $2n$ constituents times $\delta (1 - \sum\limits_{i=1}^{2n}
Z_i$). In QGSM, where both valence and sea quarks have (at the energies considered
here), the same $Z^{-1/2}$ behaviour, one has~ 

$$\rho^n(Z_1, \cdots Z_{2n}) = C_n \ Z_1^{-1/2} \ Z_2^{-1/2} \cdots Z_{2n-1}^{-1/2} \
Z_{2n}^{1.5} \ \delta \left ( 1 - \sum_{i=1}^{2n} Z_i \right )$$

\noindent where $Z_{2n}$ corresponds to the diquark and the constant $C_n$ is
determined from the nor\-ma\-li\-za\-tion to unity. In this case one gets by integrating
over all variables except the valence quark one

$$\widetilde{\rho}_{q_v}^n(y) \equiv Z \ \rho_{q_v}^n(Z) = C_n \ Z^{1/2} (1 -
Z)^{n+1/2} \ \ \ . \eqno(6)$$

\noindent In DPM, on the other hand, one assumes a $\bar{Z}^{-1}$ behaviour for the sea
quark with $\bar{Z} = \sqrt{Z^2 + \mu^2/s}$. This only changes the effective power of
$(1 - Z)$ in (6) and we have checked that at present energies the differences are only
of a few per cent. \par  

The quark momentum distribution in (6) is also appropriate for the diquark breaking case
when the latter takes place via the color exchange mechanism described in ref. \cite{[3]}.
Indeed, in this case the diquark, while changing its color, retains essentially its
momentum. However, in the case of diquark breaking via longitudinal momentum transfer,
where the two quarks in the diquark are separated, the number of independent constituents
is now $2n + 1$ (see figs. 2 and 4 for the cases $n = 1$ and $n = 2$ respectively).
Moreover, the $X$-value of one of the quarks has to be
larger than $Z$ - i.e. the momentum fraction of the quark carrying the baryon number.
\par

One then has instead of (6)

$$\widetilde{\rho}_{q_v}^n(y) \equiv Z \ \rho_{q_v}^n(Z) = C'_n \ Z^{1/2} \int_Z^{1-Z}
{dX \over \sqrt{X}} (1 - X - Z)^{n - {1 \over 2}} \ \ \ . \eqno(7)$$

\noindent In the numerical calculations we have used eq. (7). We have checked that using
(6) the changes in our results are at most of the order of $10 \ \%$. \par

For the constant $C$ in eq. (1) we take, in the case of a $NN$ collision, $C$ = 0.84.
The factor 0.84 originates as follows. In $NN$ collision we produce a net baryon
number equal to 2 with equal number of protons and neutrons. Moreover, at the
present energies we know that the number of $\Lambda/\Sigma^0$ is about 0.1 \cite{[10]}, while
the number $\sum^{\pm}$ is about 0.06 \cite{[7]}. For the diquark splitting component of
figs.~2 and 4, however, the production of strange baryons is twice as large as in the case
of direct diquark fragmentation since the $s$-quark can be created in either side of
the valence quark. (In the component depicted in fig.~5 the production of
strange baryons would be enhanced by a factor 3). Therefore from baryon number
conservation the average multiplicity of either protons and neutrons will be 1 - 0.1 -
0.06 = 0.84 (remember that $\widetilde{\rho}_{q_v}^n(y)$ is normalized to unity). \par

Finally we take $\sigma_{in} =$ 32 mb and from ref. \cite{[3]} $\sigma_{DB} =$
7 mb\footnote{Actually, two values of $\sigma_{DB}$, dependent on the form of the proton
wave function, are quoted in \cite{[3]}.
The lower one, describes well the experimental data on the energy dependence of
$(dN/dy)^p - (dN/dy)^{\bar{p}}$ at $y^* \sim 0$, (line II in fig. 3 of that
reference). This line is obtained from our eq. (1), when weighted by
$\sigma_{DB}/\sigma_{in}$ with $\sigma_{DB} \sim 6 \div 7$ mb. In the notations of ref.
\cite{[3]} this line corresponds to $\sigma_{DB} = 3.5 \ {\rm mb}/K$ with $K \sim 0.5$. In the
following we shall use the value $\sigma_{DB}$ = 7 mb. (Using $\sigma_{DB}$ = 6 mb the
changes in our results are less than 5 $\%$).}. \par

We are now ready to compute the nucleon rapidity distribution in $NN$ and $AB$
collisions. We concentrate on the difference $\Delta p = p - \bar{p}$ which has
been measured by the NA35 collaboration in peripheral and central $SS$ collisions
\cite{[7]}. It is given by

$${dN^{NN \to \Delta p} \over dy}(y) = \left ( 1 - {\sigma_{DB} \over \sigma_{in}}
\right ) {dN_{DP}^{NN \to \Delta p} \over dy}(y) + {\sigma_{DB} \over \sigma_{in}}
{dN_{DB}^{NN \to \Delta p} \over dy}(y) \ \ \ . \eqno(8)$$

\noindent Here $dN_{DB}/dy$ is given by eqs.~(1) and (7), with $n = 1$, while
$dN_{DP}/dy$ has been computed in ref. \cite{[4]} (all formulae and numerical constants are
the ones given there). \par

Likewise in an $AA$ collision we have

$${dN^{AA \to \Delta p} \over dy}(y) = \left ( \bar{n}_A - \bar{n}_A^{DB} \right ) 
\left [ {dN_{DP}^{NN \to \Delta p} \over dy}(y) \right ]_{n = \bar{n}/\bar{n}_A} + \bar{n}^{DB}_{A} \left [
{dN_{DB}^{NN \to \Delta p} \over dy} (y) \right ]_{n = \bar{n}/\bar{n}_A} \ \ . \eqno(9)$$

\noindent where
$$\left . \bar{n}_A^{DB}(b) = A \left ( 1 - \sigma_{DB} \  T_{AA}(b) \right
)^A \right / \sigma_{AA}^{DB}(b)$$

\noindent and $\bar{n}_A$ has the same expression except that $\sigma_{DB}$ is
replaced by $\sigma_{in}$ - both in the numerator and in the denominator. 
Eq. (9) is easily generalized for the case of collision of different nuclei A and B.
Distribution function $dN_{DB}/dy$ in (9) is
given by eqs.~(1) and (7) with $n = \bar{n}/\bar{n}_A$, i.e. the average number of
collisions per participant nucleon. Again $dN_{DP}/dy$ is the one computed in ref. \cite{[4]}.
For central collisions, $\bar{n}_A$ and $\bar{n}_A^{DB}$ in (9) have to be computed for
$b \sim 0$. \par

Note that by construction eqs.~(8) and (9) reproduce the rapidity distributions of
ref. \cite{[4]} for $\sigma_{DB} = 0$. The extra nucleon stopping for $\sigma_{DB} \not= 0$
is due to the fact that $\bar{n}_A^{DB}$ increases much faster than $\bar{n}_A$
with increasing $A$ - and to the different shapes of $DP$ and $DB$ rapidity
distributions. The former has a pronounced dip at $y^* \sim 0$, while the latter is
concentrated at mid rapidities. \par

The results for peripheral and central $SS$ collisions at CERN energies are shown in
fig. 6 and compared with the data from the experiment NA35
\cite{[7]}. We also show the results
obtained with $\sigma_{DB} = 0$ \cite{[4]}. We see that the dramatic dip present in the latter
case for central $SS$ collision 
has been largely filled in by the contribution of the $DB$ component in agreement
with experiment. The prediction for a central $Pb$-$Pb$ collision is also shown. In
this case the dip is converted into a broad plateau. \par

\begin{figure}[tbh]
\includegraphics{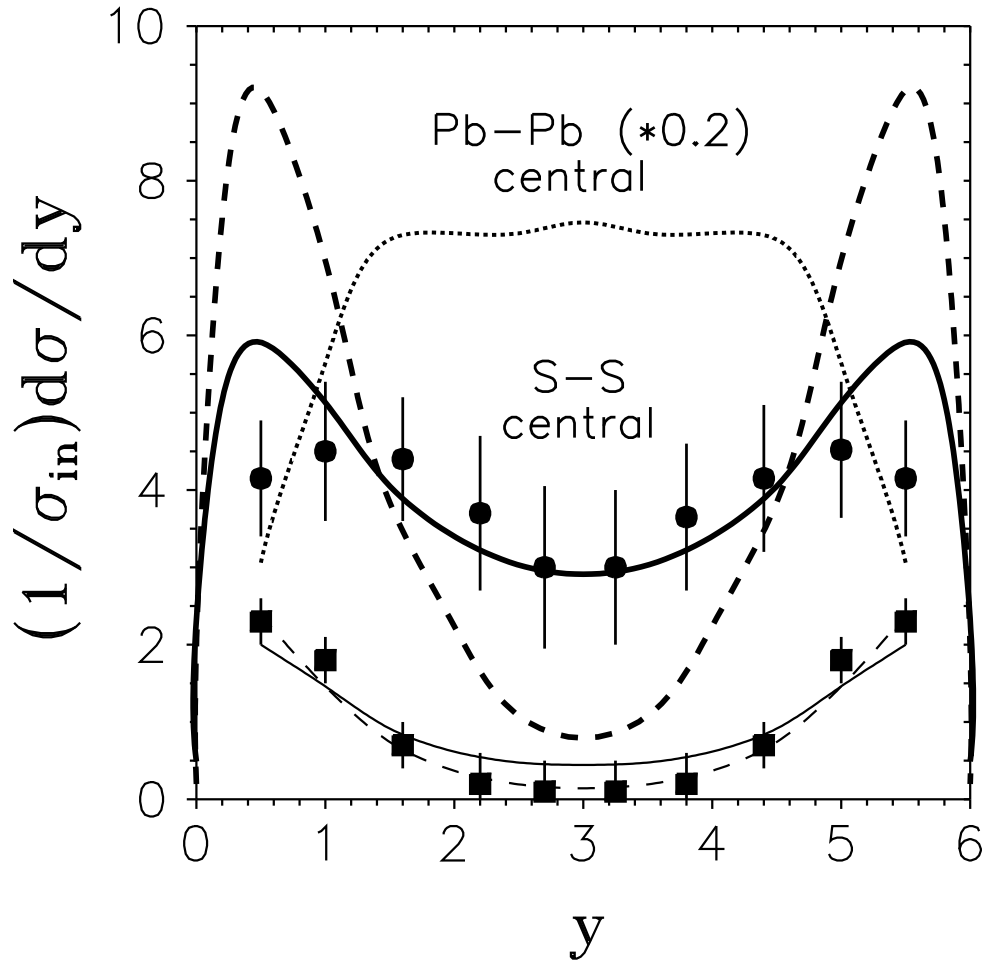}
\begin{center}
\vspace{13cm}
\parbox{13cm}
{\caption[Delta]
{Rapidity distribution of the $p$-$\bar{p}$ 
difference in AA collisions at $\sqrt{s}=20\ GeV$.
The thick dashed curve corresponds to the diquark preserving mechanism
of ref. [4] for central SS collision. 
The thick solid curve shows the result of our calculations including 
the diquark breaking component for central SS collision. The full circles
are the data points from [7]. 

The same for peripheral SS collisions is shown by
the thin solid and dashed curves and the square points.
Here the theoretical curves are for $NN$ 
collisions, normalized to the data.

The dotted curve shows our prediction for 
central $Pb$-$Pb$ collision scaled down by a factor $1/5$.
It corresponds to nucleon minus
antinucleon - rather than to proton minus antiproton.}
\label{fig6}}
\end{center}
\end{figure}

We conclude with two remarks on strange baryon production and on the extrapolation of
our results to much higher energies. Concerning the first, we have already explained
that the ratio of strange over non-strange baryons is twice as large in the $DB$
component as in the $DP$ one~: the corresponding $\Lambda$ enhancement is, however,
rather small. We obtain an extra $\Lambda$ average multiplicity of 1.1 in central $SS$
and 13 in central $Pb$-$Pb$ (with extra $y^* = 0$ densities of 0.3 and 3.8,
respectively). As discussed in ref. \cite{[4]} the existing data require final state
interactions such as $\pi + N \to K + \Lambda$. The results obtained with the nucleon
rapidity distributions computed above are close to the ones obtained in ref.~\cite{[4]},
where some extra stopping had been arbitrarily introduced (see full lines in fig.~3 of
ref.~\cite{[4]}). More precisely the $\Lambda$ distribution in central $SS$ is increased by about
10~$\%$ as compared to the one given in \cite{[4]}, and the one for central $Pb$-$Pb$
collisions is increased by about 15~$\%$ at $y^* \sim 0$ but is narrower than the one
given in \cite{[4]}. \par

Finally, we turn to the $DB$ component in fig.~5 which has not been considered so far.
As emphasized in ref.~\cite{[3]}, in this component the baryon number follows the distribution
of a sea quark or gluon and would lead to a baryon distribution flat in rapidity and
independent of energy at very high energies. At present CERN energies, however, such a
component would produce essentially no change in our results. Indeed, as discussed
above, a sea quark has an $\bar{Z}^{-1}$ distribution in DPM, and we have checked
numerically that at the energies considered here this distribution differs by less
than 10~$\%$ from the one of a valence quark. However, at energies of RHIC and especially
LHC this component would provide an interesting and
efficient mechanism of nucleon stopping. As discussed above it would also be efficient
in producing strange baryons at mid-rapidities. Actually two extreme scenarios may be
considered~: 1) the component depicted in fig.~5 is strongly suppressed for 
some dynamical
reason and can be neglected and 2) at RHIC energies and beyond, this component is the
dominant part of $DB$ - and is independent of $s$. In both cases the $DP$
component at $y^* \sim 0$, which falls roughly as $s^{-1}$, is already negligible at
RHIC energies. As for the $DB$ one, it falls in the first scenario as $s^{-1/4}$ giving a
$y^* \sim 0$ net baryon number, $p$-$\bar{p}$, of 10 at RHIC and 2 at LHC. In the second
scenario, one gets a net baryon number at $y^* \sim 0$ of 10 at both RHIC and LHC, which
is about $30 \ \%$ of the one predicted at 160 GeV/c (see fig.~6).  

Finalizing this work we learned that D.~Kharzeev has just published paper \cite{kh}
on a similar subject, and we would like to compare our results. 
The very observation of \cite{kh} that gluons can trace baryon number
was already done in \cite{[3]}, and in addition the cross section was evaluated. 
We disagree with \cite{kh}, however, on the rapidity 
dependence for the gluonic mechanism of baryon number flow. 
Solid arguments presented in \cite{kz1,kz2} (see also \cite{gn}) 
and in the present paper prove 
that this mechanism is rapidity independent. 
However, this point is not important for phenomenology at present energies
since the data \cite{[5],[6],[7]}
are dominated by the preasymptotic mechanism of a valence quark
exchange accompanied with the diquark destruction 
(exchange of a valence quark together with
the string junction in terms of topological expansion \cite{rv}).
Finally, in contrast to \cite{kh}, we have presented a full calculation
of baryon stopping in $AA$ collisions. 

\vskip 5 truemm
\noindent {\bf Acknowledgements.} \par \nobreak One of the authors (A.C.) would like to
thank D.~Kharzeev for an interesting discussion that prompted the authors to resume the
present work. He also thanks A.~Krzywicki and J.~Tran~Thanh Van for discussions. One of
the authors (B.K.) would like to thank E.~Predazzi for stimulating discussions and 
the director of the LPTHE Michel Fontannaz for the warm hospitality
extended to him at Orsay, where this work was started.

\end{document}